\documentclass[aip,jcp,showpacs,reprint]{revtex4-1}

\usepackage{CJK}
\usepackage[utf8]{inputenc}
\usepackage{amsmath}
\usepackage{graphicx}
\usepackage{dcolumn}
\usepackage{color}
\usepackage{textcomp}
\usepackage[squaren, Gray, cdot]{SIunits}
\begin{document}
\begin{CJK*}{GB}{}
%\newcolumntype{d}[1]{D{.}{.}{#1}}

\newcommand{\GN}[1]{\textcolor{magenta}{{\bf jf:} #1}}
\newcommand{\NM}[1]{\textcolor{blue}{{\bf NM:} #1}}
\newcommand{\PP}[1]{\textcolor{red}{{\bf PP:} #1}}
\newcommand{\LKB}[1]{\textcolor{green}{{\bf LKB:} #1}}
\newcommand{\JFJ}[1]{\textcolor{yellow}{{\bf JFJ:} #1}}
\newcommand{\PBRO}[1]{\textcolor{cyan}{{\bf PBro:} #1}}

\title{ Probing potential energy surface exploration strategies for complex systems }

\author{Gawonou Kokou N'Tsouaglo}

\author{Laurent Karim B\'eland}
\author{Jean-Fran\c{c}ois Joly}

\affiliation{D\'epartement de physique and Regroupement qu\'eb\'ecois sur les mat\'eriaux de pointe,
Universit\'e de Montr\'eal, C.P. 6128, Succursale Centre-Ville,
Montr\'eal, H3C~3J7, Qu\'ebec, Canada.}

\author{Peter Brommer}
\affiliation{Department of Physics and Centre for Scientific
  Computing, University of Warwick, Gibbet Hill Road, Coventry CV4
 7AL, United Kingdom}

\author{Normand Mousseau}

\affiliation{Laboratoire de Physique Th\'eorique de la Mati\`ere Condens\'ee, Universit\'e Pierre et Marie Curie, Boite 121, 4, Place Jussieu, 75252 Paris Cedex 05, France}

\affiliation{D\'epartement de physique and Regroupement qu\'eb\'ecois sur les mat\'eriaux de pointe,
Universit\'e de Montr\'eal, C.P. 6128, Succursale Centre-Ville,
Montr\'eal, H3C~3J7, Qu\'ebec, Canada.\footnote{Permanent address.}}

\author{Pascal Pochet}
\affiliation{Universit\'e de Grenoble Alpes, INAC-SP2M, L$\_{Sim}$, F-38000 Grenoble, France}
\affiliation{CEA, INAC-SP2M, Atomistic Simulation Laboratory, F-38000 Grenoble, France}

\date{\today}

\begin{abstract}

The efficiency of minimum-energy configuration searching algorithms is closely linked to the energy landscape structure of complex systems. Here we characterize this structure by following the time evolution of two systems, vacancy aggregation in Fe and energy relaxation in ion-bombarded c-Si, using the kinetic Activation-Relaxation Technique (k-ART), an off-lattice kinetic Monte Carlo (KMC) method, and the well-known Bell-Evans-Polanyi (BEP) principle. We also compare the efficiency of two methods for handling non-diffusive flickering states -- an exact solution and a Tabu-like approach that blocks already visited states. Comparing these various simulations allow us to confirm that the BEP principle does not hold for complex system since forward and reverse energy barriers are completely uncorrelated. This means that following the lowest available energy barrier, even after removing the flickering states, leads to rapid trapping: relaxing complex systems requires  crossing high-energy barriers in order to access new energy basins, in agreement with the recently proposed replenish-and-relax model [B\'eland et al., PRL \textbf{111}, 105502 (2013)] This can be done by forcing the system through these barriers with Tabu-like methods. Interestingly, we find that following the fundamental kinetics of a system, though standard KMC approach, is at least as efficient as these brute-force methods while providing the correct kinetics information.

\end{abstract}

%\pacs{61.72.-y,68.35.Gy,61.43.Bn,81.65.Mq }

\maketitle
\end{CJK*}
\section{Introduction} % (fold)
\label{sec:introduction}

Finding pathways towards global minima on the energy landscape of complex materials is a
major challenge in many fields\cite{Wales2003}. In the last decades, we have assisted to
the multiplications of new approaches for accelerating the exploration of the energy
landscape space while still attempting to follow physical-relevant pathways (see, for
example, Refs.~\onlinecite{Barkema1996,wales1997global,voter1997hyperdynamics,Goedecker2004,
beland2011}).

Because the complexity of energy landscapes increases at least exponentially with system
size\cite{Wales:1999fk, rossi2009searching}, much efforts have gone into identifying
local features that could be used to bias the search towards global low-energy
structures. Such knowledge would allow one to generate physically relevant and efficient
moves much more quickly, reducing the size of the effective landscape and increasing the
probability of constructing pathways leading to global energy minima.

Among the various propositions, a number of groups have suggested that the
Bell-Evans-Polanyi (BEP) principle \cite{Evans1935,Bell1936,Marcus1968}, developed in
physical chemistry, could also apply to more complex systems
\cite{Wales:1999fk,Goedecker2004,Roy2008,Kushima2009}. The BEP principle states that the
lowest-energy barriers surrounding a local minimum lead to deeper low-energy states;
following systematically the lowest-energy barrier out of a local minimum should
therefore rapidly lead to deep minima. It is closely connected to the methods that follow
the lowest vibrational normal mode(s) to establish folding pathways and find native
states of proteins and other molecules\cite{Bahar2005}. While the BEP principle has been
used mostly for molecules\cite{Jensen1999}, its application to bulk matter is relatively
new\cite{Goedecker2004}. Recent results, however, raise questions about its
efficiency, even in a relatively simple bulk system\cite{brommer2012comment,koziatek2013}, in apparent contradiction  with a detailed analysis proposed by Goedecker and collaborators\cite{Roy2008}. 

In this article, we review this contradiction, assessing the efficiency gains and
insights provided by the BEP principle by comparing its application with kinetic
Monte-Carlo (KMC), an algorithm known to provide the correct kinetics\cite{Bortz1975}. This comparison allows us to better understand how these methods work but, more importantly, what the nature of the energy landscape of complex bulk systems is. 

Irrespective of the sampling method, handling low-energy non-diffusive events, that would
otherwise trap the system in local energy minima, is essential. This is why the
application of most of these energy landscape exploration methods includes generally an
additional step for handling flickers, i.e.\ non-diffusive states separated by low-energy
barriers that increase the energy landscape complexity without contributing to the system
evolution, and frequently visited sates. A number of approaches have been proposed for
handling these states, including the exact treatment of their
kinetics\cite{Novotny1995,Athenes1997, puchala2010energy,
beland2011,Fichthorn:2013:164104,Cao2014} and Tabu-like methods, approaches that block
already visited states or transition, facilitating the overall phase-space
sampling\cite{Glover:1997,Chubynsky:2006aa,El-Mellouhi2008}. As it turns out, the choice
of a method for handling flickers can have major influence on the overall efficiency of a
sampling algorithm.

Here, we combine the BEP-based and the KMC strategies with both exact
flicker-handling and Tabu-like methods in order to separate the role of each of these
algorithms. To do so, we use the kinetic
Activation-Relaxation Technique (k-ART), an off-lattice KMC method with on-the-fly
catalog building, that handles both disordered systems and long-ranges deformations
directly\cite{El-Mellouhi2008,beland2011}.

In the following section we describe the implementation of the various methods. We then
present results from tests run on two systems: vacancy aggregation in iron and
relaxation of an ion-implanted box of crystalline silicon. The signification of these
results is presented in the discussion section. When handling flicker states correctly,
we find that crossing high-energy barriers is essential to open new low-energy pathways,
by moving into unvisited energy basins that can lead to new low-energy structures. On the
other hand, while Tabu does not preserve the correct kinetics, it significantly raises
the efficiency of BEP, but does not significantly accelerate the configurational space
sampling as compared with standard KMC with flicker-handling.

\section{Methodology} % (fold)
\label{sec:methodology}

The comparison presented in this paper is done between two algorithms, kinetic Monte Carlo and the BEP principle.  For each of these methods, we apply two different approaches for handling low-energy barriers. All these are run using the kinetic Activation-Relaxation Technique (k-ART) package as a base.  In this section, we first describe the k-ART package and then each algorithm separately. 

\begin{figure}[t]
		\includegraphics[height=2.5in]{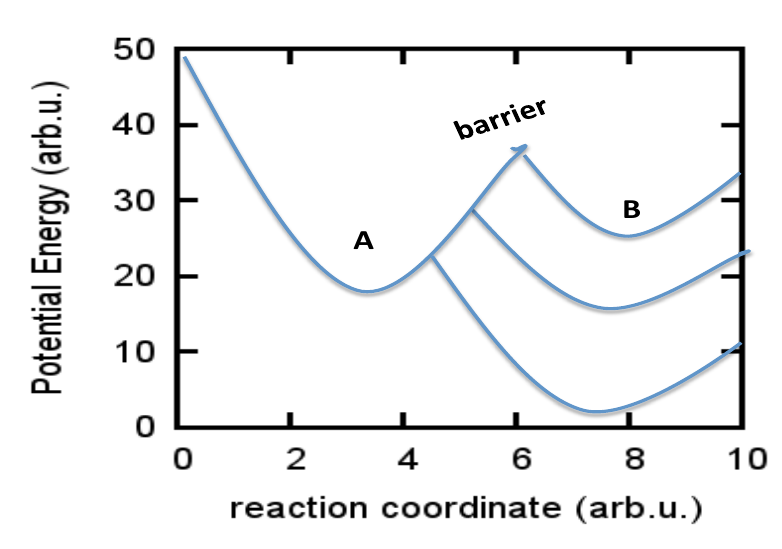}
	\caption{The Bell-Evans principle: if all local-energy basins are similar in size, then selecting the lowest-energy barrier from an initial minimum (A) will lead to the lowest-energy minimun in (B) and an overall faster energy relaxation.}
	\label{fig:fig1}
\end{figure}

\subsection{The kinetic Activation-Relaxation Technique} % (fold)
\label{sub:the_kinetic_activation_relaxation_technique}

The kinetic Activation-Relaxation Technique (k-ART) is a kinetic Monte Carlo algorithm
(KMC)\cite{Bortz1975} that lifts many of the technical restrictions preventing its
application to complex materials\cite{El-Mellouhi2008,beland2011}. Traditionally, KMC
uses a fixed, preconstructed event catalog to compute the rate of escape from a local
minimum and brings forward the simulation clock according to a Poisson
distribution.\cite{Bortz1975,fichthorn1991theoretical}. This choice limits the atomic
motion to discrete states, which are generally crystalline positions, preventing its
application to disordered or defective materials, alloys and, in many cases,
semiconductors, and leaving aside much of long-range elastic effects on energy barriers
and kinetics.

While k-ART is described in details in Refs.~\onlinecite{beland2011, Joly2012,Mousseau2012}, it is useful to provide here a short description of the algorithm. Updating the system in k-ART can be described as a four-step process:
\begin{enumerate}

\item After a move, all atoms are inspected for changes in local environment. A spherical region around each atom, with a radius typically set to between 5 and 7~\AA, is defined. A bonding graph is constructed between atoms within this region, by connecting nearby atoms within a preset cut-off, generally fixed between the first and second-neighbor. Using NAUTY\cite{McKay1981}, a topological analysis library, we identify the unique automorphic group associated with this bonding graph, irrespective of the various symmetry operations. This allows us to construct a discrete and reusable catalog even for totally disordered systems.

\item For each new topology encountered, excluding the crystalline ones that would only lead to improbable events on the simulation timescale, we launch a series of event searches using the latest version of ART nouveau\cite{Malek2000, Machado-Charry2011}. For the two systems studied here, we launch 50 random event searches for each new topology and restrict our search to events with an energy barrier less or equal to 5~eV, generating on average between 3 and 5 events per topology and therefore per atom in a non-perfectly crystalline environment.  To ensure a complete catalog, new searches are also regularly launched on the most common topologies. 

\item Once the catalog is updated to include events associated with the new topologies, all events corresponding to the current configuration are placed in a binary tree in preparation for the KMC step. All barriers corresponding to at least 99.99\% of the rate, computed with constant prefactor, are then reconstructed and fully relaxed to account for geometrical rearrangements due to short- and long-range elastic deformations. The final individual and global rates are therefore associated with the exact conformation. 

\item Finally, the standard KMC algorithm is applied to select an event to execute it and
advance the clock according to a Poisson distribution. Once the event is executed, we return to (1) for the next step.

\end{enumerate}

Using topological classification coupled with unbiased open search for transition states,
k-ART handles events without regard to the presence or not of a crystalline substructure,
constructing the event-catalog as the system evolves and fully taking care of all elastic
effects. Parallelizing event searches over tens to hundreds of processors  \cite{joly2012,beland2011}, K-ART has been applied with success to highly defective crystals\cite{brommer2012comment,
beland2013replenish,beland2013long,brommer2014}, alloys and even amorphous materials\cite{joly2013contribution},
generating atomistic trajectories on time scale of 1~s or more and providing insight in
the long-time dynamics of these systems.

\subsection{Implementing the Bell-Evans-Polanyi Principle} % (fold)
\label{sub:implementing_the_bell_evans_polanyi_principle}

The Bell-Evans-Polanyi (BEP) principle is based on the observation that the local curvature on the energy landscape is almost constant for given systems\cite{Roy2008}. Taken to its extreme (see Fig.~\ref{fig:fig1}), BEP implies that the barrier height out of a local minimum is a direct indicator of the depth of the following energy minimum. To relax a system, it is therefore sufficient to always select the lowest available energy-barrier.

Implementing the Bell-Evans-Polanyi principle is straightforward within k-ART. We follow
steps (1) to (3) according to the description above. The only difference is that after all
barriers have been relaxed, the lowest-energy barrier is systematically selected within the limits of the flicker handling method as discussed in the next subsection. Although time has no physical meaning with the BEP approach, we still use the KMC rate to assign a clock to the BEP evolution for comparison with k-ART results.

\subsection{Handling flickers} % (fold)
\label{sub:handling_the_flickers}

The efficiency of event-based simulations is limited by the presence of flickers,
non-diffusive states of similar energy separated by a low-energy barrier with respect to
those leading to structural evolution. When the KMC or BEP strategy, as defined above, is
applied to any system with more than a few barriers, simulations  become
trapped rapidly within flickers that seize all computational efforts without structural
evolution. Many efforts have gone into handling flickers since KMC was first introduced
to material sciences, 25 years ago\cite{Bortz1975}. Here we consider two approaches:
the basin auto-constructing Mean-Rate Method (bac-MRM)\cite{beland2011}, that we have
adapted from Puchala \emph{et al.}'s Mean-Rate Method\cite{puchala2010energy} and a
simple barrier-based Tabu\cite{Glover:1997,Chubynsky:2006aa,El-Mellouhi2008}.

The bac-MRM, discussed in details in Ref.~\onlinecite{beland2011}, handles flickers by merging the associated states within a single basin,
solving the internal dynamics analytically, projecting the solution onto the various exit
pathways, and correcting their respective rate. Since the bac-MRM is statistically exact
and since by definition the in-basin states have very close energy, it is possible to
adjust the basin barrier cut-off as the simulation evolves to prevent it from being trapped.

When the focus is on sampling configurations rather than follow the right kinetics, it is
possible to limit or even forbid the visit of already known states. In barrier-based
Tabu, when a barrier is selected, we compare the trajectory, i.e.\ the displacement from
the initial to saddle to final state, with the last $N$ moves (see 
Ref.~\onlinecite{Chubynsky:2006aa} for more details). If the displacement is not in the database, the
event is generated, otherwise, the configuration is left in the initial or final state
according to their respective Boltzmann weight. The transition can be completely
forbidden of the rest of the simulation or blocked for a number $n$ of steps, hence the
name \emph{Tabu}. Here we select $n=50$.

While recent k-ART simulations with the KMC algorithm used the bac-MRM, which
offers a statistically correct kinetics\cite{beland2011}, the BEP implementation of
Goedecker's basin-hoping minimization algorithm\cite{Goedecker2004} is closer in spirit
to the Tabu approach than to bac-MRM. Indeed, in basin-hopping, while states are never
blocked, the energy landscape exploration is controlled by adaptable energy pulses. The
energy of these pulses is raised systematically as the same basins are visiting multiple
times so that, as for Tabu and contrary to bac-MRM, there is no formal upper limit to the
energy barrier that can be crossed when the system is trapped in a local basin.

\subsection{Systems studied} % (fold)
\label{sub:systems_studied}

We compare the KMC and BEP methods and the impact of flicker handling on two
different systems: (1) the aggregation of 50 vacancies inside a 2000-atom box of bcc iron
described with the Ackland  potential \cite{ackland2004development} and (2) the
relaxation of a 27000-atom box of c-Si disordered through the implantation of a
single 3-keV Si ion and described with Stillinger-Weber potential\cite{stillinger1985computer}. Both systems are run at
300 K and at constant volume corresponding to crystalline density.

For the Iron system, we start with a 2000-atom bcc Fe cubic box and remove 50 atoms at random. Both BEP and KMC
simulations are launched after a simple local energy minimisation. For the Ackland 
potential, the vacancy diffusion barrier is found to be 0.64 eV with
MD\cite{ackland2004development} and ART nouveau. At 300 K, aggregation from
random vacancies into 9 to 10 vacancy clusters was found to take on the order of 1~ms in
two three independent off-lattice KMC simulations\cite{brommer2012comment,xu2013cascade,chill2014molecular}.

The initial configuration of ion-implanted Si is described in detail in
Ref.~\onlinecite{beland2013replenish}. A 3-keV ion is first implanted in a 100~000-atom
Stillinger-Weber box\cite{stillinger1985computer} with two surfaces along the
z-direction and periodic boundary condition (PBC) along the $x$ and $y$ directions and is
then relaxed for 10~ns using NVT molecular dynamics at 300~K. A block of 27~000 atoms
surrounding the disordered region is then extracted and placed into a cubic cell with PBC
along the three axes. The kinetics of relaxation with k-ART and bac-MRM was found to be
in excellent agreement with nanocalorimetric measurements\cite{beland2013replenish, beland2013long}.

\section{Results} % (fold)
\label{sec:results}

\subsection{Basin Mean Rate Method} % (fold)
\label{sub:basin}
\subsubsection{ Vacancies in bcc iron}

\begin{figure}[t]
		\includegraphics[height=2.2in]{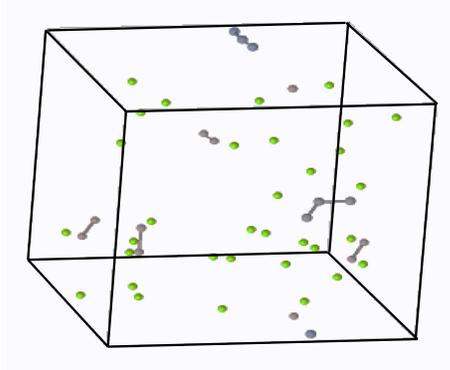}
	\caption{ The initial state for a 2000-atom bcc-iron box containing 50 vacancies colored as function of cluster size. Mono-vacancies are colored in green, cluster containing two vacancies are colored in grey and tri vacancies are colored in dark.} 
	\label{fig:fig3}
\end{figure}

\begin{figure}[t]
		\includegraphics[height=2.5in]{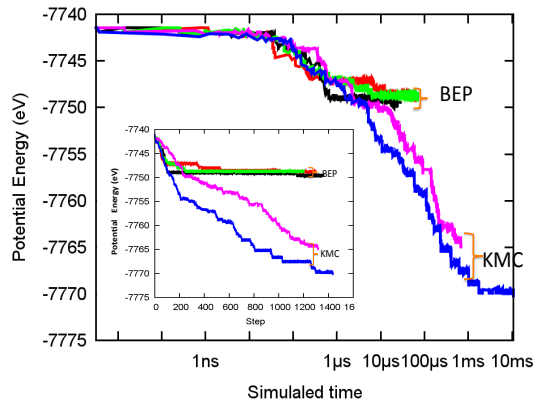}
	\caption{ The total energy evolution as function of logarithmic of simulated time for three BEP and two  KMC  runs in the 50-vacancy Fe system.. Inset: Total energy evolution as function of simulation step for three BEP and two KMC runs.}
	\label{fig:fig4}
\end{figure}

We first compare the k-ART and BEP relaxation methods coupled with the bac-MRM using the Fe system. Fig.~\ref{fig:fig3} shows the initial state for a 2000-atom bcc-iron box
containing 50 vacancies colored as function of cluster size. We run three  independent simulations for BEP and two for KMC. Each run is about 1300 k-ART steps not counting flickering steps that are handled with the bac-MRM. Fig.~\ref{fig:fig4} reports the evolution of the total energy as a function of time (as a function of step in inset).

The five simulations, using either k-ART or BEP, follow a similar trajectory for the
first 100 steps or so. At that point, all BEP simulations are trapped at an energy about
7~eV below the initial configuration for the rest of the simulations (more than 1000
steps further for each run), unable to find pathways to more relaxed states while the
KMC simulations evolve the system for the whole run, finishing
between 25 and 28 eV below the BEP runs. Projecting these runs on a time axis, we see
that the two methods follow the same path until about \unit{10}{\micro}s, at which point
the clock for BEP runs slows down noticeably compared to KMC: after 1300 steps, BEP runs
reach about \unit{100}{\micro}s compared with 1 to 10 ms for KMC.

\begin{figure}[t]
		\includegraphics[height=2.5in]{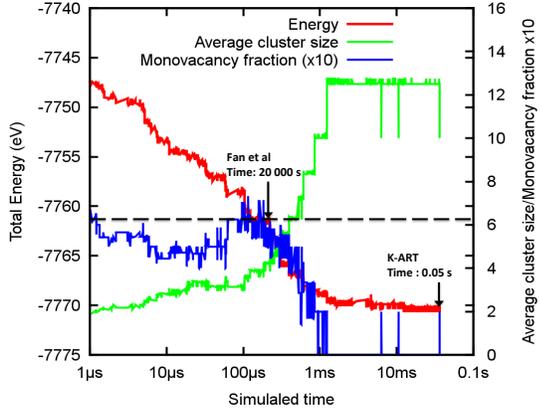}
	\caption{Comparison of k-ART vs. ABC relaxation of an 2000-atom iron system with 50 vacancies initially distributed at random. Red line: total energy evolution as function of logarithmic of simulated time; green line: evolution of cluster size; blue line: evolution of the fraction of mono-vacancies. The horizontal dashed black line corresponds to the energy level reached after 20~000 s with ABC.}
	\label{fig:fig5}
\end{figure}

This difference in effective time is not caused by the handling of flickers, since both k-ART and BEP use, here, the bac-MRM. Indeed, these  BEP simulation results are consistent with Fan \emph{et al.}\cite{Fan:2011ab} recent work using the Autonomous Basin Climbing (ABC) method, a BEP-like approach\cite{Kushima2009}. Using the same 50-vacancies Fe system, ABC simulations produced an energy  drop of 13~eV during a simulation lasting 20~000~s, while  k-ART reaches the same energy level in the first 500~\textmu s of simulation and continues to relax well-below ABC's level. Fig.~\ref{fig:fig5}  (red line) compares the performance of k-ART with KMC with that of Autonomous Basin Climbing (ABC) for this system.

\begin{figure}[t]
		\includegraphics[height=2.8in]{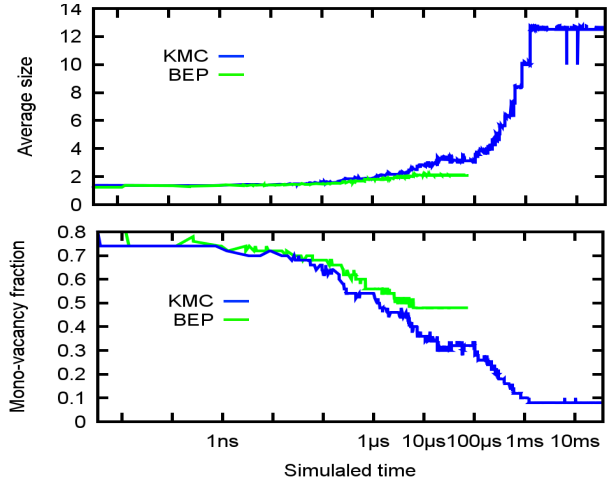}
	\caption{Comparison of k-ART vs.\ BEP with structural evolution for the 50-vacancy FE system. Top: Evolution of the average cluster size as function of logarithmic of
simulated time. Bottom: Evolution of the fraction of mono-vacancies. Blue line: KMC; green line: BEP.}
	\label{fig:fig6}
\end{figure}

\begin{figure}[t]
		\includegraphics[height=2.8in]{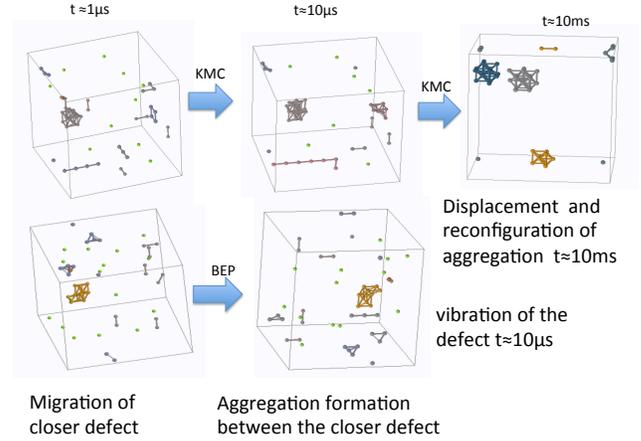}
	\caption{Selected snapshots of the atomic configuration for a KMC run (top) at 1~\textmu s, 10~\textmu s and 10~ms and a BEP run at 1~\textmu s and 10~\textmu s. Only vacancies are shown. Colors are associated with cluster size. Mono-vacancies are coloured in green, cluster containing two vacancies are colored in grey and tri vacancies are colored in dark.}
	\label{fig:fig7}
\end{figure}

To understand the difference between these two methods, we look at the time evolution of the average vacancy cluster and the mono-vacancy fraction for one BEP and KMC simulation (Fig.~\ref{fig:fig6}). These correspond respectively to  the green and  blue lines in Fig.~\ref{fig:fig4}.  As for the energy, structural evolution for the two simulation types follows a similar path for the first 10 \textmu s, which corresponds to the clustering of about 38~\% of the initial value of the vacancies into small clusters (averaged cluster size equals two). At that stage, the structural evolution of the BEP run comes almost to a stop while the aggregation continues with KMC simulation with clusters reaching an average size of 13 as the proportion of mono-vacancies falls to less than 12.5~\% of the initial value. This supports the relation between BEP and Fan \emph{et al.} simulation observed for the total energy (blue line in Fig.~\ref{fig:fig5}), where the mono-vacancy fraction decreased to only 52~\% of the initial value (averaged size 6) after 20~000~s.  The structural difference between the final BEP and KMC states is clearly seen in the snapshots taken during the evolution of both simulation types (Fig.~\ref{fig:fig7}). Even at 10 \textmu s, we note a difference in the number of isolated vacancies between the two types of runs. 
 
\begin{figure}[t]
		\includegraphics[height=2.5in]{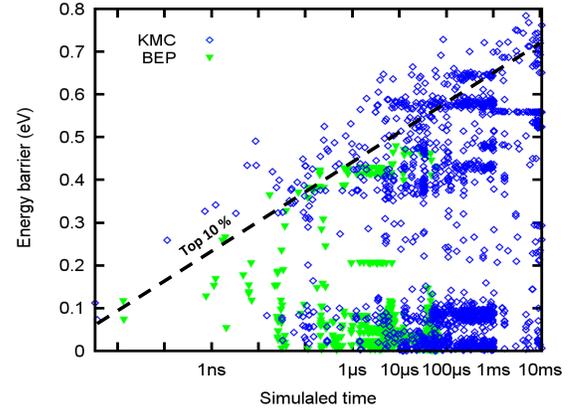}
	\caption{Distribution of the executed energy barriers as function of the logarithmic simulated time for both BEP (green triangle) and KMC (blue squares) simulations of the Fe 50-vacancy system. The dashed black line provides a rough estimate of the energy threshold for the events with barrier height in the top 10~\% of the executed events in a given time interval.}
	\label{fig:fig8}
\end{figure}

To further understand the kinetic evolution of these two simulations sets and their relation to the structure of the energy landscape, we analyze
the evolution of the energy barrier height for all \emph{executed} events.
Fig.~\ref{fig:fig8} shows all the energy barriers for executed events as function of
logarithmic time of KMC and BEP simulations. For KMC simulations, we note that the
maximum barrier height increases logarithmically with time, but that, in any time frame, the energy barrier distribution remains almost continuous.  For BEP simulations
we note instead, that by 10 ns, the maximum barrier height visited -- around 0.4~eV -- has
been reached and that, after this point, the same distribution of barrier is selected
until the simulations stopped, around 10 \textmu s, after 1300 steps. KMC manages therefore
to access, even at times as short as 10~ns, activated barriers that are slightly
higher, 0.5~eV, than those crossed with BEP, but sufficient to unlock configurations by giving access to new relaxation pathways.

\begin{figure}[t]
		\includegraphics[height=2.5in]{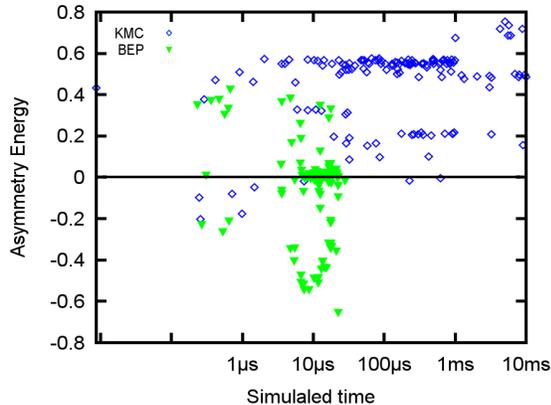}
	\caption{Asymmetric energy, i.e.\ energy difference between the final and initial energy states, for events within the top highest 10~\% energy barrier as function of logarithmic simulated time in the Fe 50-vacancy system. (93~\% of these events lead in high energy state in KMC simulations vs 53~\% events in BEP simulations)}
	\label{fig:fig9}
\end{figure}

Why would crossing high barriers be so important? Fig.~\ref{fig:fig9} plots the energy released by the system, or the asymmetry energy, for executed events associated with the 10~\% higher energy barriers calculated in a moving window. In this plot, negative asymmetry energy means that the system has moved into a state of lower energy while positive values are associated with higher energy final states. We see that 93~\% of these events lead to states with a higher final energy  in KMC simulation versus 53~\% events in BEP simulation. In BEP, crossing these high energy barriers leads, almost half the time, to lower energy states and, as often, to higher energy states while, for KMC runs, the bias is clearly towards higher energy states. 

Before discussing the significance of this observation, we first need to check whether these results are seen in other systems.

\subsubsection{ Ion-bombarded crystalline silicon}

\begin{figure}[t]
		\includegraphics[height=2.5in]{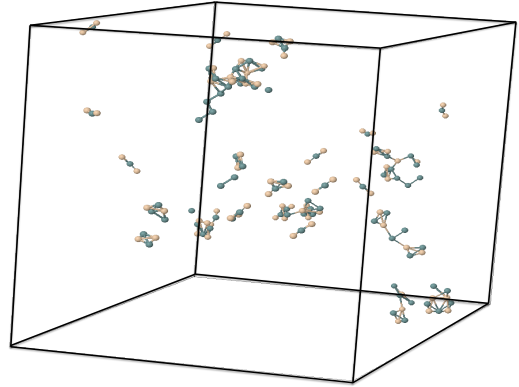}
	\caption{The initial state for a 27000-atom box of c-Si after the implantation of a single 3~keV Si atom and relaxation, through MD, over 1~ns. Interstitials are colored in beige and vacancies in blue.}
	\label{fig:fig10}
\end{figure}

We consider now a disordered system with
an equal number of vacancies and interstitials: ion-implanted crystalline silicon. This
model system has been studied extensively, both experimentally and theoretically, over the
years\cite{caturla1996ion,karmouch2007damage,pothier2011flowing, beland2013replenish}. Here, we follow the relaxation of a 27000-atom box of Stillinger-Weber
\cite{stillinger1985computer} c-Si disordered through the implantation
 of a single 3-keV Si. After 1~ns of MD, the implantation
yields 152 defects (interstitials and vacancies) distributed into about 30 clusters
ranging in size from 1 to 30 with most of them counting between 4 and 7 defects. This configuration serves as the starting point for all simulations in this section. A
snapshot of this initial configuration is presented in Fig.~\ref{fig:fig10}. 

\begin{figure}[t]
		\includegraphics[height=2.5in]{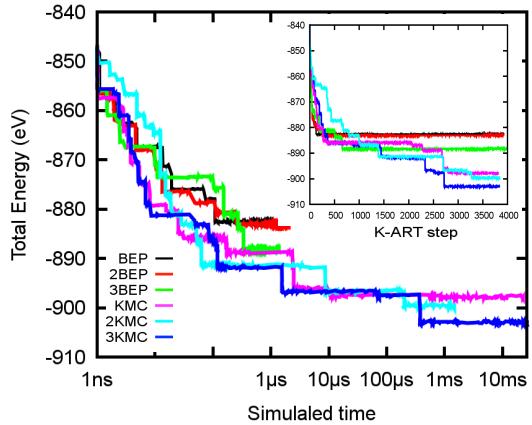}
	\caption{The total energy evolution as function of the logarithm of simulated time for 3  KMC and 3 BEP runs on a 27~000-atom cell of ion-implanted Si. The k-ART runs are launched after a first 1~ns molecular dynamics following a 3~keV single Si implantation. Inset: Total energy evolution as function of simulation step for the same runs.}
	\label{fig:fig11}
 \end{figure}

Fig.~\ref{fig:fig11} presents the energy evolution of 3 BEP and 3 KMC simulations over time and, in inset, as a function of k-ART steps. All simulations are run for about 4000 k-ART steps with an overall similar behavior for BEP and KMC runs to what had been observed for vacancy aggregation in Fe: both BEP and KMC sets follow each other closely for the first \textmu s, corresponding to about 1200 k-ART steps. After this point, the BEP remains trapped onto a constant energy surface while the KMC runs release another 25~eV and access a time scale of up to 10 ms.

The difference in relaxation pathway between BEP and KMC simulations is analyzed using
the defect evolution. In the first 100 ns, for both systems, we observe only cluster reconfigurations and single defect migration.  Thus, the number of defects remains constant (152). 
Afterwards, annihilation events dominate during \unit{1}{\micro}s and the number of defects passes from 152 to 88 where
BEP simulations is locked, vibrating and reconfiguring the defects while k-ART with KMC
simulations continue to generate annihilation events by effectively moving distant
defects, allowing a further reduction in their number to 64.

\begin{figure}[t]
		\includegraphics[height=2.5in]{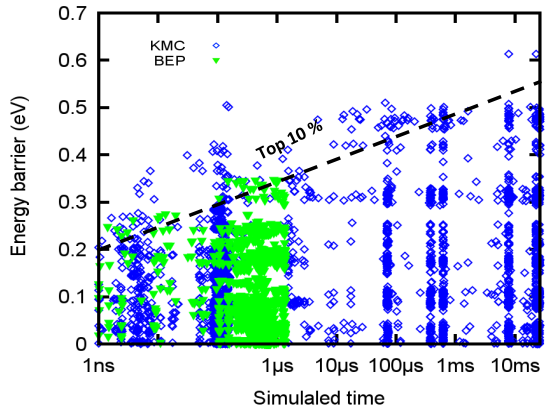}
	\caption{Distribution of the selected energy barriers as function of the logarithmic simulated time for both BEP (green triangle) and KMC (blue squares) ion implanted Si simulations of a 27~000-atom cell of ion-implanted Si. The black line provides a rough estimate of energy threshold for the events with barrier height in the top 10~\% of the selected events in a given time interval.}
	\label{fig:fig13}
 \end{figure}

\begin{figure}[t]
		\includegraphics[height=2.5in]{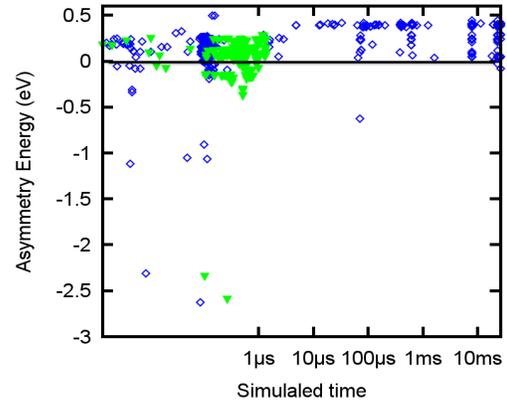}
	\caption{Asymmetry energy, i.e.\ energy difference between the final and initial energy states, for events within the top highest 10~\% energy barrier as function of logarithmic simulated time for all BEP and KMC runs on the 27~000-atom implanted Si cell. Blue symbols: KMC runs; green symbols: BEP.  (The 89~\% events lead in high energy state in KMC simulation vs 43~\% events in BEP simulation.)}
	\label{fig:fig14}
 \end{figure}

Analysis of the energy barrier height evolution for all \emph{executed} events for the
three BEP and three KMC simulation shows the same characteristic as observed previously
in the Fe system (Fig.~\ref{fig:fig13}). Fig.~\ref{fig:fig14} gives the energy released
by the system for the top 10\% highest executed energy barriers crossed calculated over a
moving time window.  89~\% of these events lead to higher final energy states in KMC simulations vs
43~\% events for BEP simulations confirming, here also, the importance of allowing the crossing of barriers that are not the lowest in order to open up new pathways leading, in the end, to lower energy states\cite{beland2013replenish}.

 \begin{figure}[t]
		\includegraphics[height=2.5in]{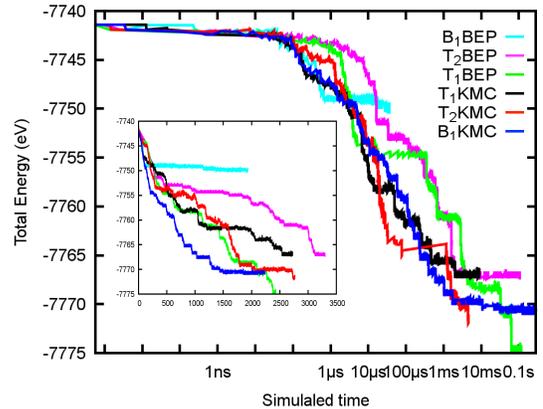}
	\caption{The total energy evolution for a system of 50 vacancies in Fe as function of logarithmic of simulated time for 2 runs of BEP with Tabu and 1 run with  bac-MRM, and 2 runs of KMC with Tabu and 1 run with  bac-MRM. Inset: Total energy their  evolution as function on simulation step.}
	\label{fig:fig15}
\end{figure}
 
 \begin{figure}[t]
		\includegraphics[height=2.5in]{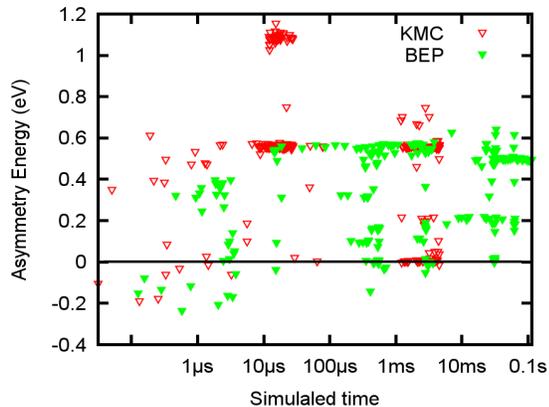}
	\caption{Asymmetric energy, i.e.\ energy difference between the final and initial energy states, for events within the top highest 10~\% energy barrier as function of logarithmic simulated time for the Tabu-based KMC and BEP simulations of the 50-vacancy Fe system.}
	\label{fig:fig17}
 \end{figure}

\subsection{Tabu} % (fold)
\label{sub:tabu}

\begin{figure}[t]
		\includegraphics[height=2.5in]{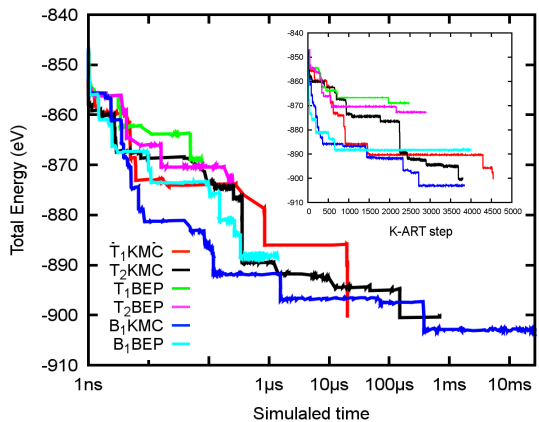}
	\caption{ The total energy evolution as function of the logarithmic simulated time for 2 runs of BEP with Tabu and 1 run with  bac-MRM, and 2 runs of KMC with Tabu and 1 run with  bac-MRM (Blue ligne in Fig.~\ref{fig:fig11}) in a 27~000-atom cell of ion-implanted Si.     Inset: Total energy their  evolution as function on simulation step.}
	\label{fig:fig18}
\end{figure}

We now look at the effect of Tabu, an approach that can be applied to handle flickers in kinetic simulations but also to orient relaxation when searching for global minimum. We compare Tabu with bac-MRM using both BEP and KMC sampling techniques.  Fig.~\ref{fig:fig15}  shows the evolution of the total energy for a 2000-atom Fe box with 50 vacancies as a function of simulation time (inset, as a function of k-ART steps). Focusing on energy as a function of time,  we note that all simulations follow a relatively similar relaxation pathway over 10 ms, with the exception of BEP-bac-MRM (light blue) that remains trapped at -7750~eV (as was previously seen). 

As a function of k-ART step (see inset), however, KMC-bac-MRM still provides the fastest overall relaxation, reaching -7770 eV after 2250 steps, almost 80 \% faster than Tabu-KMC or BEP. Nevertheless, in the long run, Tabu, irrespective of the sampling method, manages to reach KMC-bac-MRM's relaxation level and even, in one simulation, achieve a better energy gain. 

Fig.~\ref{fig:fig17} gives the energy released for all \emph{selected} events for one
Tabu-BEP and one Tabu-KMC simulation for the top 10\% highest executed energy barriers
crossed calculated over a moving time window. Analysis of the energy barrier height
evolution for these simulations shows that Tabu-based simulations display a similar rate
of visiting higher final energy states as previously observed for KMC-bar-MRM: 84~\% of
Tabu-KMC events and 93~\% Tabu-BEP events lead to higher energy states. This explains why
Tabu approaches can be, on average, as efficient as KMC-bac-MRM for finding low-energy
states. By blocking already visited directions, Tabu effectively forces the system to
sample the more asymmetric states that lead to overall lower-energy configurations. This might explain also why, since minima-hopping uses a Tabu-like approach, by systematically increased the exit energy, the method remains efficient even though it is based on the BEP principle\cite{Goedecker2004,Roy2008}.

For large scale complex system Tabu become less efficient(Fig.~\ref{fig:fig18}). The
total energy evolution as function of logarithmic of simulated time of a 27000-atom box
of c-Si disordered through the implantation of a single 3-keV Si ion for two runs of BEP
with Tabu and one run with bac-MRM, and two runs of KMC with Tabu and one run with
bac-MRM (inset: total energy their evolution as function on simulation step). We see the
two simulations using Tabu with BEP held at high energy state and Tabu with KMC
simulation descending following the same pathway as with bac-MRM with KMC , due, in part,
to the 50-step memory used here .
 
In spite of these similarities, it is important to remember that  bac-MRM is statistically exact, contrary to Tabu, and that it preserves the correct dynamics of the system. For example, analysis of energy evolution for the four Tabu simulations shows  sudden staircase-like decrease for the energy (Fig.~\ref{fig:fig18}), a behavior that is not observed with the bac-MRM simulations.

% subsection tabu (end)

% section results_and_discussion (end)

\section{Discussion} % (fold)
\label{sec:discussion}

In the previous section, we have presented four sets of simulations, each performed on
two systems: a 2000-atom Fe cell with 50 vacancies and a 30~000-atom Si cell disordered
by a 3-keV implanted atom. The results allow us to better understand the applicability of the BEP principle to bulk systems, the importance of the flicker-handling methods as well as more about the structure of the energy landscapes. 

As discussed in the introduction, the BEP principle was formulated many decades ago and it has helped to understand various kinetic phenomena in chemistry. In its simplest form, it states that around a given local energy minimum, the lowest energy-barrier will lead to the lowest energy state among those directly connected to the initial minimum. In a recent paper, Roy and collaborators showed that a relaxed version of BEP is applicable to bulk systems: crossing a small barriers has more chances to lead to a deep minimum that crossing a high-energy barrier.\cite{Roy2008} 

This observation is correct, but incomplete. Extensive characterization of the energy
landscape of amorphous silicon (a-Si), for example, has shown that in fact for any event
the forward and reverse barrier height, i.e.\ the barrier height computed from the
initial and the final minima, respectively, are totally uncorrelated,\cite{Kallel2010}.
This general observation that also holds, at least, in ion-bombarded
Si\cite{beland2013long}. Since the depth of the final well, as measured from the initial
state, is the difference between the reverse and the forward barrier height, this absence
of correlation means that, on average, lower forward energy barriers do lead to deeper
minima, which explains some of the success of the application of the principle to
materials\cite{goedecker2005global, kazachenko2009improved}. Fundamentally, however, the
structure of the landscape does not correspond to the original BEP principle, which
states that there is a direct correlation between the barrier height and the depth of the
final minimum. 

Since the BEP principle is correct locally, on average, even though for the wrong
reasons, is strictly following the lowest available energy barrier, in respect of the BEP
principle, an efficient global minimization approach? 

Results presented in the previous section show that it is not. Indeed, for two very different systems, a comparison of the BEP principle with an off-lattice kinetic Monte-Carlo, an algorithm that provides the correct atomistic kinetics, show that simulations following the BEP principle rapidly become trapped in relatively high-energy configurations while KMC simulations, that select events with a Boltzmann weight on the activation barriers, allowing them to sample beyond the lowest-energy barrier, manage to visit much deeper energy basins for the same computational effort. 

To relax efficiently, both the Fe vacancy and the ion-implanted Si systems require therefore to cross barriers that do not correspond to the lowest ones available in order to land into higher energy states, in contradiction with the locally-derived BEP model. As was shown recently in ion-implanted c-Si and a-S, accessing these high-energy states is essential to open new low-energy pathways, by moving into unvisited energy basins that can lead to low-energy structures\cite{beland2013replenish, joly2013contribution}. This so-called \emph{replenish and relax} mechanism explains why, when treating correctly the local flickering dynamics, BEP approaches cannot be as efficient as standard KMC methods. Clearly, systematically selected lower energy barriers are not sufficient to exit local energy basins.

Why  then do BEP-based relaxation methods, such as minima-hopping work? The crucial step is in the way the code handles local-energy traps. In the second set of simulations presented above, we compare the efficiency of two algorithms designed specifically to take care of the flickering states, these non-diffusive states separated by low-energy barriers with respect to those necessary to reach new energy basins. 

Here, we looked at the basin auto-constructing Mean-Rate Method, an algorithm that statistically solves the inner-basin kinetics, based on a pre-defined maximum barrier height for defining flickers, and a Tabu-like approach, that blocks already-visited transition, forcing the system to jump over higher and higher barriers until it finds a new energy basin. When coupled to KMC, both methods show a similar efficiency in finding low-energy structures, with slight advantage for the correct kinetics on the more complex ion-implanted Si system.

When applied to BEP runs, Tabu manages to prevent the trajectories to get trapped,  allowing the simulations to reach energy level similar to those obtained with the KMC method, in a similar number of steps. This is done, essentially, by violating the BEP principle, and systematically blocking the lowest energy barrier, allowing the system to cross over higher energy ones. 

These results confirm that efficient energy minimization in a bulk system cannot be based on the BEP principle. This failure is due, in part, to the fact that this principle is not exact: it is the absence of correlation between the forward and backward energy barriers that explains why lower-energy barriers tend to lead to deeper energy minima, not a specific correlation between these two quantities. More important, however, is the need to go over higher-energy states in order to reach new deep-energy minima. This can only be done either by using a physically-based kinetic algorithm such as simple MD or KMC or by systematically limiting the available phase space to non-visited regions, in effect forcing the system to move over these high-energy barriers.

% section discussion (end)

\section{Conclusion} % (fold)
\label{sec:conclusion}

In this paper, we use kinetic ART, an efficient on-the-fly off-lattice kinetic Monte Carlo algorithm that incorporates exactly all elastic effects to compare both  the Bell-Evans-Polanyi (BEP) principle with KMC and a Tabu-like approach to handling small flickering-states with a statistically exact one. 

Testing these methods on two systems --- a 2000-atom cell of EAM bcc iron with 50 vacancies and a 27~000-atom ion-bombarded crystalline silicon cell --- we find that pure BEP simulations, even when handling low-energy flickering states, become trapped rapidly in relatively high-energy configurations while KMC runs manage to find ever lower energy states (on the simulation time scale). This is explained by the \emph{replenish and relax} model\cite{beland2013replenish} which shows that to move into a new and deeper energy basin, it is necessary, first, to move into higher energy states that are not allowed by the BEP approach but that come out naturally from a kinetically-based sampling method such as KMC. It is possible to overcome BEP's limits by adding a Tabu criterion on the visited transition states. Even a relatively short memory kernel, with 50 states, is sufficient to bring the efficiency of the BEP method on par with KMC, even though the correct kinetics is lost. 

This comparison of various algorithms used for sampling energy landscape allows us to better understand the crucial \emph{replenish and relax} steps, necessary to escape local energy minima in complex system, confirming recent results\cite{beland2013long} and helping to understand the strength of kinetically-based methods for relaxing these structures.

% section conclusion (end)

\begin{acknowledgments}

  This work has been supported by the Canada Research Chairs program
  and by grants from the Natural Sciences and Engineering Research
  Council of Canada (NSERC) and the \textit{Fonds Qu\'eb\'ecois de la
   Recherche sur la Nature et les Technologies} (FQRNT). We are
  grateful to \textit{Calcul Qu\'ebec} (CQ) for generous allocations of computer
  resources. Gawonou Kokou N'Tsouaglo acknowledges financial support from 
  Islamic Development bank (IDB)  %\FE{RQCHP - superseded by CQ}

\end{acknowledgments}

%\bibliography{bep.bib}


\end{document}